\documentclass[a4paper]{jpconf}
\usepackage{graphicx}

\usepackage[colorlinks=true,linktocpage=true,linkcolor=blue,citecolor=blue]{hyperref}
\usepackage[usenames,dvipsnames]{color}

\begin{document}
\title{New formulation of leading order anisotropic hydrodynamics}

\author{LeonardoTinti}

\address{Institute of Physics, Jan Kochanowski University, PL-25406~Kielce, Poland}

\ead{tinti@fi.infn.it}

\begin{abstract}

Anisotropic hydrodynamics is a reorganization of the relativistic hydrodynamics expansion, with the leading order already containing substantial momentum-space anisotropies. The latter are a cause of concern in the traditional viscous hydrodynamics, since large momentum anisotropies generated in ultrarelativistic heavy-ion collisions are not consistent with the hypothesis of small deviations from an isotropic background, i.e., from the local equilibrium distribution.

We discuss the leading order of the expansion, presenting a new formulation for the (1+1)--dimensional case, namely, for the longitudinally boost invariant and cylindrically symmetric flow. This new approach is consistent with the well established framework of Israel and Stewart in the close to equilibrium limit (where we expect viscous hydrodynamics to work well). If we consider the (0+1)--dimensional case, that is, transversally homogeneous and longitudinally boost invariant flow,  {the new form of anisotropic hydrodynamics leads to better agreement with known solutions} of the Boltzmann equation than the previous formulations, especially when we consider finite mass particles. 

\end{abstract}

\section{Introduction}

Relativistic hydrodynamics plays a fundamental role in modeling of relativistic heavy-ion collisions, see for instance Refs.~\cite{Israel:1979wp,
Muronga:2003ta,
Baier:2006um,
Romatschke:2007mq,
Dusling:2007gi,
Luzum:2008cw,
Song:2008hj,
Denicol:2010tr,
Schenke:2011tv,
Shen:2011,
Bozek:2011wa,
Niemi:2011ix,
Bozek:2012qs,
Denicol:2012cn}. Early calculations were based on perfect fluid hydrodynamics, however, nowadays viscous codes are preferred. Both because they provide a better description of the data and because of general arguments that the fluid viscosity cannot be zero, which follows from quantum mechanical considerations~\cite{micro} as well as from the AdS/CFT correspondence~\cite{ADS/CFT}. Despite its obvious success, there are still fundamental issues with the ordinary viscous hydrodynamics expansion. A new approach to treat these problems is {\it anisotropic hydrodynamics} (aHydro)~\cite{Florkowski:2010cf,Martinez:2010sc,
Ryblewski:2010bs,Martinez:2010sd,
Ryblewski:2011aq,Martinez:2012tu,
Ryblewski:2012rr,Ryblewski:2013jsa,
Florkowski:2012ax,Florkowski:2012as,
Florkowski:2011jg,Florkowski:2014txa,
Florkowski:2014sfa}, where the large momentum anisotropy, providing large pressure corrections, is treated in a non perturbative way starting from the leading order of the hydrodynamics expasion. 

\section{The hydrodynamics expansion}

\begin{figure*}[t]
\begin{center}
\includegraphics[angle=0,width=0.75\textwidth]{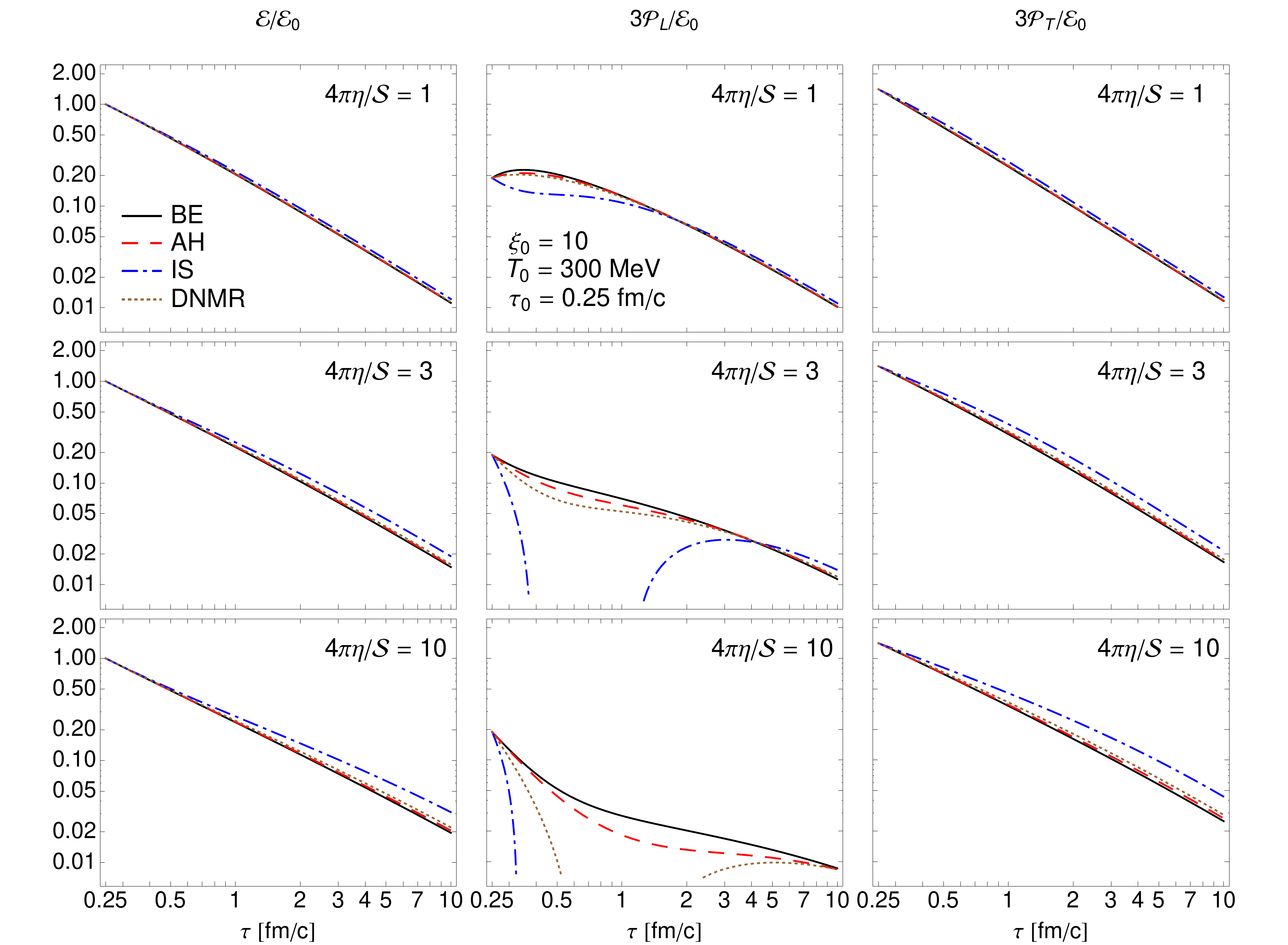}
\end{center}
\caption{(Color online) Comparison of viscous hydro with anisotropic hydrodynamics and second-order viscous hydrodynamics (figure taken from \cite{Florkowski:2013lya}).
}
\label{fig:PLPTE_300_10}
\end{figure*}

The most common assumption for deriving hydrodynamics from relativistic kinetic theory is that the particle distribution function $f(x,p)$ is very close to local equilibrium. Ignoring conserved charges and in the Boltzmann limit we have
\begin{equation}
 f(x,p) = f_{\rm eq.}(x,p) + \delta f(x,p), \qquad f_{\rm eq.}(x,p) = k \exp\left[-\frac{p\cdot U(x)}{T(x)} \right],
\label{hydro_expansion}
\end{equation}
with $T$ and $U^\mu$ being the effective temperature and the fluid four velocity, respectively. The leading order in~(\ref{hydro_expansion}), $f_{\rm eq.}$, describes the perfect fluid. The viscous correction depends only on $\delta f$ which is treated as a small perturbation. However, when we consider an (almost) boost invariant flow like the one we expect in the early stages of heavy ions collisions, we encounter fundamental problems. The four velocity gradients are inversely proportional to the proper time, therefore, the pressure corrections become close to the equilibrium pressure, questioning the validity of the perturbative treatment.

The main feature of anisotropic hydrodynamics is to treat the large momentum anisotropy in a non perturbative way starting from the leading order, namely, we write

\begin{equation}
 f(x,p) = f_{\rm aniso.}(x,p) +\delta \tilde{f}(x,p).
\label{aniso_exp}
\end{equation}
In this way, the deviation $\delta\tilde{f}$ from the (non isotropic and dissipative) background $f_{\rm aniso.}$ can be small enough to justify a perturbative treatment. The first formulation of aHydro used  the point dependent version of the Romatschke-Strickland form (presented in~\cite{Romatschke:2003ms}) for the leading order of the anisotropic expansion, which in the local rest frame (LRF) reads

\begin{equation}
 f_{\rm aniso.}(x,p) = k\exp\left[ -\frac{1}{\Lambda(x)}\sqrt{ \frac{}{} p^2_T +\zeta(x) p_L^2} \right].
\label{RS}
\end{equation}
Here $\Lambda$  is the momentum scale (the effective temperature $T$ is defined using the Landau matching and is different from $\Lambda$ in general), $p_T$ and $p_L$ are the transverse and longitudinal momenta, and $\zeta$ is the anisotropy parameter. In order to close the system of equations for the leading order of the anisotropic expansion, one has used the four momentum conservation (the first moment of the Boltzmann equation and the Landau matching) and the particle creation equation (the zeroth moment of the Boltzmann equation). In addition, the collisional kernel was treated in the relaxation time approximation.

For a longitudinally boost invariant and transversely homogeneous system there is an exact solution of the relativistic Boltzmann equation~\cite{Florkowski:2013lya}. We show in Fig.~\ref{fig:PLPTE_300_10} one of the plots in~\cite{Florkowski:2013lya}. The comparison is done between the exact solution (BE), Israel-Stewart theory, the new formulation of second-order viscous hydrodynamics presented in~\cite{Denicol:2012cn}, and anisotropic hydrodynamics (AH). Anisotropic hydrodynamics is always very close to the exact solution, while IS is providing unphysical vanishing longitudinal pressure ${\cal P}_L$, and significant deviations from the exact evolution of the temperature $T$ and the transverse pressure ${\cal P}_T$. In the most extreme case, even DNMR approach is not reliable.

\section{New formulation of the leading order}

\begin{figure}[ht]
\begin{minipage}[b]{0.37 \linewidth}
\centering
\includegraphics[angle=0,width=1.1 \textwidth]{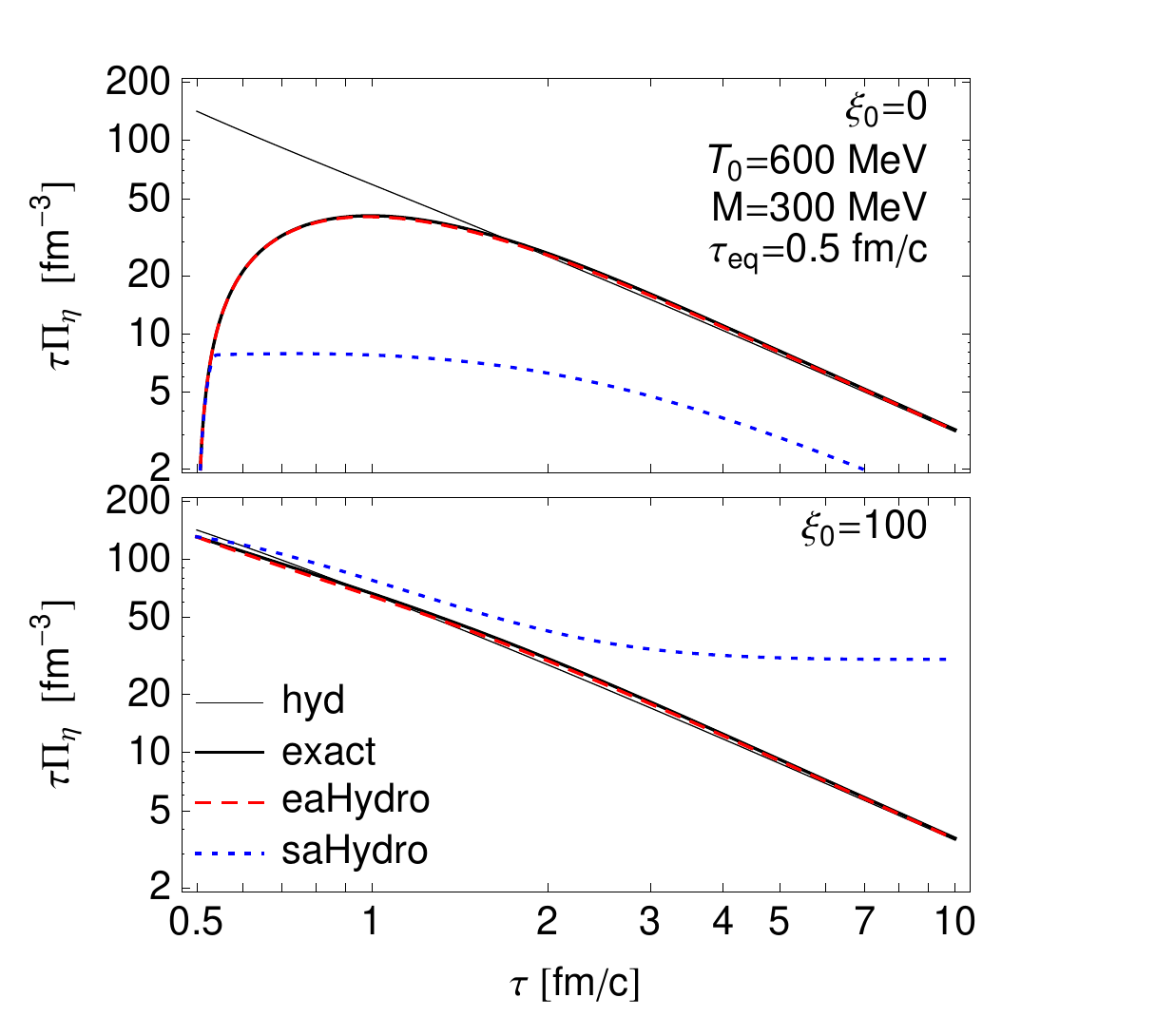}
\end{minipage}
\hspace{1.0cm}
\begin{minipage}[b]{0.37\linewidth}
\centering
\includegraphics[angle=0,width=1.1 \textwidth]{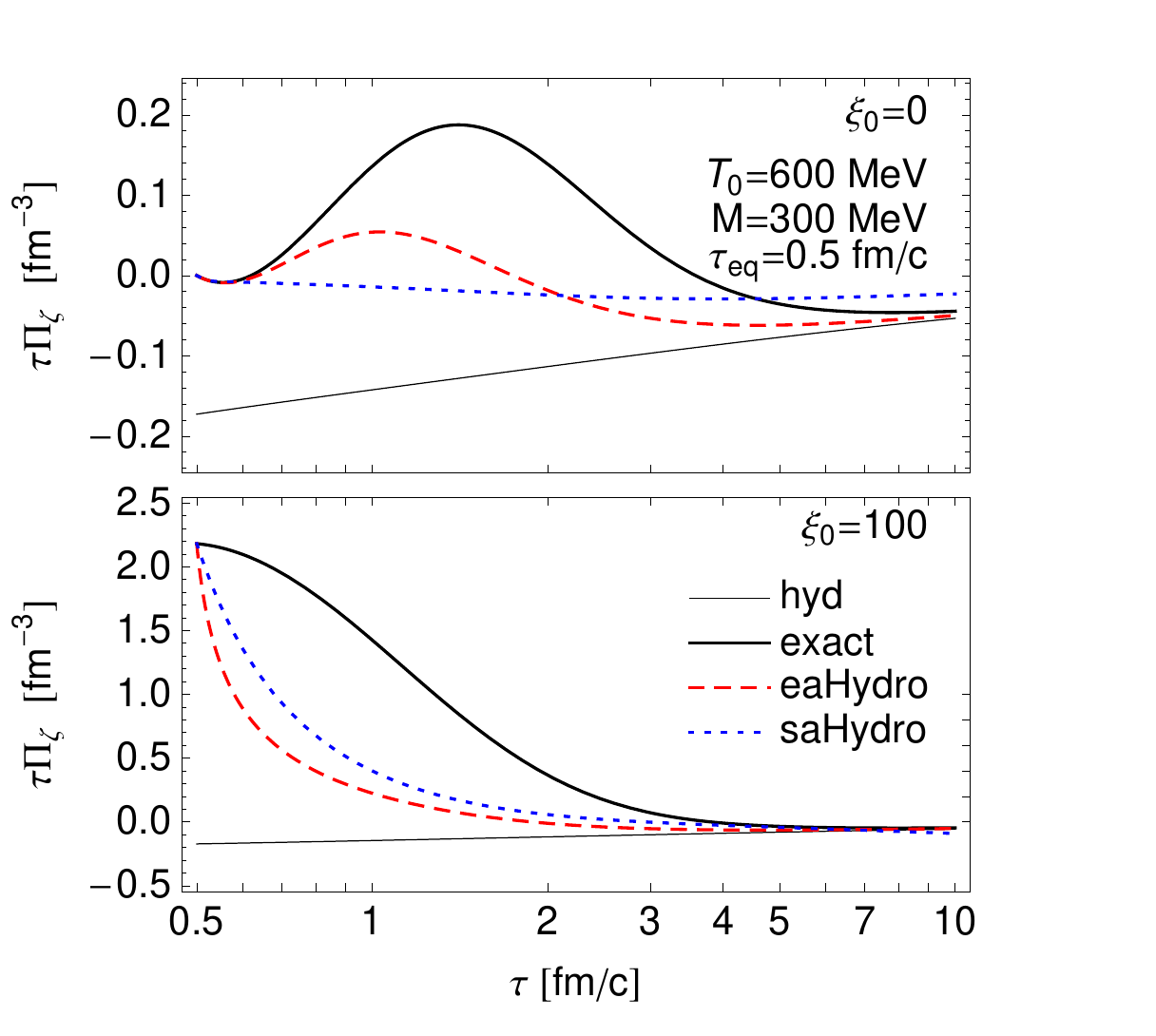}
\end{minipage}
\caption{(Color online) Time dependence of shear and bulk viscous pressure multiplied by $\tau$ (figure taken from ~\cite{Florkowski:2014nc}).}
\label{figs:shear_bulk}
\end{figure}

The anisotropic background~(\ref{RS}) takes into account differences between the longitudinal pressure ${\cal P}_L$ and the transverse pressure ${\cal P}_T$, only. However if there is a non-vanishing radial flow we expect anisotropies even in the transverse plane. As the system evolves toward equilibrium, these corrections become more important. One way to handle non trivial transverse dynamics is to treat $\delta\tilde{f}$ in the anisotropic expansion~(\ref{aniso_exp}) in a perturbative way~\cite{Bazow:2013ifa}. Alternatively,  we propose here to include more dynamic effects connected with anisotropy in the leading order itself.

In Ref.~\cite{Tinti:2014conf} we extended the formalism of anisotropic hydrodynamics to the (1+1)--dimensional case. We started from a generalization of the Romatschke-Strickland form, which in the local rest frame reads
\begin{equation}
 f_{\rm aniso.} (x,p )= k \exp\left[ -\frac{1}{\lambda(x)}\sqrt{ \frac{}{} (1+\xi_X)p_X^2 + (1+\xi_y)p_Y^2 + (1+\xi_Z)p_Z^2 } \right],
\end{equation}
where $Z$ is the longitudinal direction, and $X$ is the direction of the transverse flow. We used the second moment of the Boltzmann equation, in addition to the energy and momentum conservation, in order to obtain a closed set of equations. We proved that these equations reduce to the Israel-Stewart equations in the close to equilibrium limit, where we know that second-order viscous hydrodynamics is justified.

We later compared this new set of equations with the solution of the Boltzmann equation and the original prescription for anisotropic hydrodynamics~\cite{Florkowski:2014nc}. There is a large improvement of the agreement with the exact solution, especially for massive particles. In Fig.~\ref{figs:shear_bulk} we show the comparison between the new formulation (eaHydro) and the original one (saHydro). The shear evolution $\tau\Pi_\eta$ is very well reproduced, while the bulk evolution $\tau\Pi_\zeta$ still shows some deviations from the exact solution. Note that $\tau$ is the (longitudinal) proper time, ${\cal P}_{\rm eq.}$ is the equilibrium pressure, $ \Pi_\eta = \frac{2}{3}\left( {\cal P}_T -\frac{1}{3}{\cal P}_L\right)$, and $\Pi_\zeta = \frac{1}{3}\left( 2{\cal P}_T + {\cal P}_L -3{\cal P}_{\rm eq.} \right)$.
%
%

\section{Conclusions}

Anisotropic hydrodynamics is a reorganization of the hydrodynamic expansion around a non-isotropic background. The leading order already provides large longitudinal pressure corrections, justifying the perturbative treatment of the next to leading order in heavy ion collisions. The original prescription for the leading order of anisotropic hydrodynamics does not take into account pressure anisotropies in the transverse plane, therefore requiring a next to leading order treatment in presence of transverse expansion. We extended the original treatment allowing for cylindrically symmetric expansion already in the leading order. The agreement with the exact solution in the case of vanishing transverse flow has been largely improved. Bulk dynamics is not well reproduced, however, an interesting proposal is to introduce an extra degree of freedom taking into account the isotropic pressure corrections, see~Ref.~\cite{Nopoush:2014nc}.

\section*{Acknowledgements}

This work has been supported by Polish National Science Center grant No. DEC-2012/06/A/ST2/00390.

\end{document}